\documentclass[aps,prl,nobalancelastpage,superscriptaddress,twocolumn,longbibliography]{revtex4-2}
\usepackage{graphicx} % Required for inserting images
\usepackage{standalone, gensymb, tensor, amsmath,amsfonts,amssymb,amsthm,bbm,bm, braket}
\usepackage[T1]{fontenc}
\usepackage{hyperref}
\usepackage[inline]{enumitem}
\usepackage{scalerel, physics}
\usepackage{wasysym}
\usepackage{comment}
\usepackage{enumerate}
\usepackage{graphicx}
\usepackage{mathtools}
\usepackage{soul}
\usepackage{xparse}
\usepackage{dsfont}%added this package for the identity symbol (Alessandro)
\usepackage{ifthen}
\usepackage{tensor}
\usepackage{tikz}
\usepackage{pgfplots}
\usepackage{tikz-network}
\usepackage{simplewick} 

\usetikzlibrary{patterns,decorations.pathreplacing}
\theoremstyle{break}        

\theoremstyle{break}

\usepackage{color}
\definecolor{OliveGreen}{RGB}{85,107,47}
\definecolor{NavyBlue}{RGB}{0,0,128}
\usepackage{tensor}
\usepackage{xifthen}
\usepackage{xargs}

\newcommandx{\fineq}[5][1=-.8ex,2=1,3=1,5=0]{%Alessandro modified, added an option to rotate the picture around the origin(can be useful sometimes)
	\begin{tikzpicture}[baseline={([yshift=#1]current  bounding  box.center)}, scale = #2, every node/.style={scale = #3},rotate around={#5:(0,0)},every node/.style={transform shape}]
		#4
	\end{tikzpicture}
}

\definecolor{bertinired}{RGB}{232,102,102}
\definecolor{bertiniblue}{RGB}{101,147,245}
\definecolor{bertinigreyblue}{RGB}{166,218,149}
\definecolor{bertinigreyred}{RGB}{232,102,102}
\definecolor{bertinivioletc}{RGB}{45,130,60}
\definecolor{bertinigreen}{RGB}{166,218,149}
\definecolor{bertiniorange}{RGB}{255, 116, 23}
\definecolor{OliveGreen}{RGB}{85,107,47}
\definecolor{NavyBlue}{RGB}{0,0,128}
\definecolor{bertiniviolet}{RGB}{210,145,178}
\definecolor{bertinigrey1}{RGB}{98,98,98}
\definecolor{bertinigrey2}{RGB}{211,211,211}
\definecolor{bertinigrey3}{RGB}{192,192,192}
\definecolor{bertinigrey4}{RGB}{169,169,169}

\newcommandx{\tikzdiagup}{
	\tikz {\draw[thick] (0,0)--(0.15,0.15); \draw (0,0) rectangle (0.15,0.15);}
}

\newcommandx{\gatecross}[1][1=0.5]{
	\pgfmathparse{#1/2.0}
	\let\x\pgfmathresult
	\draw[thick] (-\x,-\x) -- (\x,\x);
	\draw[thick] (\x,-\x) -- (-\x,\x);
}
\newcommandx{\gatesqu}[2][1=0.25,2=]{
	\pgfmathparse{#1/2.0}
	\let\x\pgfmathresult
	\ifthenelse{\equal{#2}{}}{
		\draw[thick, fill=white, rounded corners=2pt] (-\x,\x) rectangle (\x,-\x);
	}{
		\draw[thick, fill=#2, rounded corners=2pt] (-\x,\x) rectangle (\x,-\x);
	}
}

\newcommandx{\gatemark}[2][1=0.075,2=tr]{
	\pgfmathparse{#1}
	\let\l\pgfmathresult
	\ifthenelse{\equal{#2}{topleft}}{
		\draw[thick] (0,\l) -- ++(-\l,0) --++ (0,-\l);
	}{}
	\ifthenelse{\equal{#2}{topright}}{
		\draw[thick] (0,\l) -- ++(\l,0) --++ (0,-\l);
	}{}
	
	\ifthenelse{\equal{#2}{bottomleft}}{
		\draw[thick] (0,-\l) -- ++(-\l,0) --++ (0,\l);
	}{}
	\ifthenelse{\equal{#2}{bottomright}}{
		\draw[thick] (0,-\l) -- ++(\l,0) --++ (0,\l);
	}{}
	
}

\newcommandx{\roundgate}[6][1=0,2=0,3=1,4=topright,5=white,6=1]{
	\pgfmathparse{#3}
	\let\l\pgfmathresult
	\begin{scope}[shift={(#1,#2)}]
		\gatecross[\l]
		\pgfmathparse{\l/2.0}
		\let\s\pgfmathresult
		\gatesqu[\s][#5]
		\pgfmathparse{\l*0.15}
		\let\m\pgfmathresult
		\ifthenelse{\equal{#6}{1}}{		\gatemark[\m][#4]
		}{	\node at ({0},{0}) {\scalebox{1.3}{$#6$}};}
	\end{scope}
}
\newcommandx{\wcirc}[2]{\begin{scope}
		\draw[fill=white] (#1,#2) circle (0.15);	\end{scope}} %Alessandro command, just to standardize the size of the circle
\newcommandx{\wcircc}[2]{\begin{scope}
		\draw[fill=white] (#1,#2) circle (0.13);	\end{scope}} %Alessandro command, just to standardize the size of the circle

\newcommandx{\wsqr}[2]{\begin{scope}
		\draw[fill=white,shift={(#1,#2)}] (-.13,.13) rectangle (.13,-.13);	\end{scope}} %Alessandro command, just to standardize the size of the circle
\newcommandx{\wsqrr}[2]{\begin{scope}
		\draw[fill=white,shift={(#1,#2)}] (-.11,.11) rectangle (.11,-.11);	\end{scope}} %Alessandro command, just to standardize the size of the circle
\newcommandx{\bcirc}[2]{\begin{scope}
		\draw[fill=black] (#1,#2) circle (0.15);	\end{scope}} %Alessandro command
\newcommandx{\thetastate}[4][1=0,2=0,3=1,4=]{
	\pgfmathparse{#3/2}
	\let\l\pgfmathresult
	\pgfmathparse{\l*0.15}
	\let\m\pgfmathresult
	\begin{scope}[shift={(#1,#2)}]
		\draw[thick] (0,0)--(\l,\l);
		\draw[thick] (0,0)--(-\l,\l);
		\ifthenelse{\equal{#4}{}}{
			\draw[fill=white] (0,0) circle (0.15);
		}{
			\draw[thick, fill=#4] (0,0) circle (0.15);
		}
	\end{scope}
}

\newcommandx{\thetastateflipped}[4][1=0,2=0,3=1,4=]{
	\pgfmathparse{#3/2}
	\let\l\pgfmathresult
	\pgfmathparse{\l*0.15}
	\let\m\pgfmathresult
	\begin{scope}[shift={(#1,#2)}]
		\draw[thick] (0,0)--(\l,-\l);
		\draw[thick] (0,0)--(-\l,-\l);
		\ifthenelse{\equal{#4}{}}{
			\draw[fill=white] (0,0) circle (0.15);
		}{
			\draw[thick, fill=#4] (0,0) circle (0.15);
		}
	\end{scope}
}

\newcommandx{\vertgate}[5][1=0,2=0,3=4,4=bertiniorange,5=topright]
{
	\begin{scope}[shift={(#1,#2)}]
		\ifthenelse{\equal{#3}{1}}{
			\roundgate[0][0][1][#5][#4]
		}{
			\foreach \n[evaluate=\n as \y using {2*\n-2}] in {1,...,#3}{
				\roundgate[0][\y][1][#5][#4]
			}
		}
	\end{scope}
}

\newcommandx{\tsfmatV}[8][1=0,2=0,3=l,4=4,5=tr,6=init,7=bertiniorange,8=topright]{
	\begin{scope}[shift={(#1,#2)}]
		\ifthenelse{\equal{#3}{l}}{
			\pgfmathsetmacro{\flag}{0}
		}{
			\pgfmathsetmacro{\flag}{1}
		}
		
		\foreach \y[evaluate=\y as \x using {mod(\y+\flag,2)}] in {1,...,#4}{
			\roundgate[\x][\y][1][#8][#7]
		}
		\ifthenelse{\equal{#5}{tr}}{
			\foreach \y[evaluate=\y as \x using {mod(\y+\flag,2)}] in {#4}{
				\draw [fill=white] (\x-0.5,\y+0.5) circle (0.15);
				\draw [fill=white] (\x+0.5,\y+0.5) circle (0.15);
			}
		}{}
		\ifthenelse{\equal{#6}{init}}{
			\thetastate[\flag][0][1][#7]
		}{}
	\end{scope}
}

\newcommandx{\leftriangle}[5][1=0,2=0,3=4,4=bertiniorange,5=topright]{
	\begin{scope}[shift={(#1,#2)}]
		\pgfmathsetmacro{\t}{#3}
		\pgfmathsetmacro{\steps}{ceil(\t/2)}
		\foreach \i[evaluate=\i as \x using -\t+2*\i-1, evaluate=\i as \ylim using \t-2*\i+2] in {1,...,\steps}{
			\foreach \y[evaluate=\y as \thisx using {\x+\y-1}] in {1,...,\ylim}{
				\roundgate[\thisx][\y][1][#5][#4]
			}
		}
	\end{scope}
}

\newcommandx{\rightriangle}[5][1=0,2=0,3=4,4=bertiniorange,5=topright]{
	\begin{scope}[shift={(#1,#2)}]
		\pgfmathsetmacro{\t}{#3}
		\pgfmathsetmacro{\steps}{ceil(\t/2)}
		\foreach \i[evaluate=\i as \x using -\t+2*\i-1, evaluate=\i as \ylim using \t-2*\i+2] in {1,...,\steps}{
			\foreach \y[evaluate=\y as \thisx using {-\x-\y+1}] in {1,...,\ylim}{
				\roundgate[\thisx][\y][1][#5][#4]
			}
		}
	\end{scope}
}

\newcommandx{\eigenVL}[8][1=0,2=0,3=l,4=5,5=tr,6=init,7=bertiniorange,8=topright]{
	\begin{scope}[shift={(#1,#2)}]
		\pgfmathsetmacro{\t}{#4}
		\leftriangle[0][0][\t][#7][#8]
		
		\ifthenelse{\equal{#6}{init}}{
			\drawinitstate[0][0][l][\t][#7]
		}{}
		
		\ifthenelse{\equal{#5}{tr}}{
			\draw[fill=white] \foreach \x in {0,...,\t} {(\x-0.5-\t,0.5+\x) circle (0.15)};
			\ifthenelse{\equal{#3}{r}}{
				\draw[fill=white] (0.5,\t+0.5) circle (0.15);
			}{}
		}{}
		\ifthenelse{\equal{#5}{parttr}}{
			\draw[fill=white] \foreach \x in {0,...,\t} {(\x-0.5-\t,0.5+\x) circle (0.15)};
		}{}
	\end{scope}
}

\newcommandx{\eigenVR}[8][1=0,2=0,3=l,4=5,5=tr,6=init,7=bertiniorange,8=topright]{
	\begin{scope}[shift={(#1,#2)}]
		\pgfmathsetmacro{\t}{#4}
		\rightriangle[0][0][\t][#7][#8]
		
		\ifthenelse{\equal{#6}{init}}{
			\drawinitstate[0][0][r][\t][#7]
		}{}
		
		\ifthenelse{\equal{#5}{tr}}{
			\draw[fill=white] \foreach \x in {0,...,\t}{(-\x+0.5+\t,0.5+\x) circle (0.15)};
			\ifthenelse{\equal{#3}{l}}{
				\draw[fill=white] (-0.5,\t+0.5) circle (0.15);
			}{}
		}{}
	\end{scope}
}

\newcommandx{\tra}[2][1]{\underset{#1}{\text{tr}}\left[#2\right]}%Alessandro command, wanted to uniformize the notation for trace/partial trace

\newcommandx{\tsfmatDgate}[7][1=0,2=0,3=l,4=4,5=tr,6=bertiniorange,7=topright]
{
	\begin{scope}[shift={(#1,#2)}]
		\ifthenelse{\equal{#3}{l}}{
			\pgfmathsetmacro{\flag}{-1}
		}{
			\pgfmathsetmacro{\flag}{1}
		}
		\pgfmathsetmacro{\t}{#4}
		\foreach \i[evaluate=\i as \x using {\flag*\i}, evaluate=\i as \y using \i] in {1,...,\t}{
			\roundgate[\x][\y][1][#7][#6]
		}
		
		\ifthenelse{\equal{#5}{tr}}{
			\foreach \i[evaluate=\i as \x using {\flag*\i}, evaluate=\i as \y using \i] in {\t}{
				\draw [fill=white] (\x-0.5,\y+0.5) circle (0.15);
				\draw [fill=white] (\x+0.5,\y+0.5) circle (0.15);
			}  
		}{}
	\end{scope}
	
}

\newcommandx{\tsfmatD}[8][1=0,2=0,3=l,4=4,5=tr,6=init,7=bertiniorange,8=topright]{
	\begin{scope}[shift={(#1,#2)}]
		\ifthenelse{\equal{#6}{init}}{
			\thetastate[0][0][1][#7]
		}{}
		
		\ifthenelse{\equal{#3}{l}}{
			\pgfmathsetmacro{\flag}{-1}
		}{
			\pgfmathsetmacro{\flag}{1}
		}
		
		\pgfmathsetmacro{\t}{#4}
		\foreach \i[evaluate=\i as \x using {\flag*\i}, evaluate=\i as \y using \i] in {1,...,\t}{
			\roundgate[\x][\y][1][#8][#7]
		}
		
		\ifthenelse{\equal{#5}{tr}}{
			\foreach \i[evaluate=\i as \x using {\flag*\i}, evaluate=\i as \y using \i] in {\t}{
				\draw [fill=white] (\x-0.5,\y+0.5) circle (0.15);
				\draw [fill=white] (\x+0.5,\y+0.5) circle (0.15);
			}  
		}
		\ifthenelse{\equal{#5}{parttr}}{
			\foreach \i[evaluate=\i as \x using {\flag*\i}, evaluate=\i as \y using \i] in {\t}{
				\draw [fill=white] (\x+0.5*\flag,\y+0.5) circle (0.15);
			}  
		}
		{}
	\end{scope}
}

\newcommandx{\drawinitstate}[5][1=0,2=0,3=l,4=4,5=bertiniorange]{
	\pgfmathsetmacro{\t}{#4}
	\begin{scope}[shift={(#1,#2)}]
		\pgfmathsetmacro{\steps}{ceil((\t-1)/2)}
		\ifthenelse{\equal{#3}{l}}{
			\foreach \i[evaluate=\i as \x using -\t+2*\i] in {0,...,\steps}{
				\thetastate[\x][0][1][#5]
			}
		}{
			\foreach \i[evaluate=\i as \x using -\t+2*\i] in {0,...,\steps}{      
				\thetastate[-\x][0][1][#5]
			}
		}
	\end{scope}
}

\newcommandx{\drawinitstateflipped}[5][1=0,2=0,3=l,4=4,5=bertiniorange]{
	\pgfmathsetmacro{\t}{#4}
	\begin{scope}[shift={(#1,#2)}]
		\pgfmathsetmacro{\steps}{ceil((\t-1)/2)}
		\ifthenelse{\equal{#3}{l}}{
			\foreach \i[evaluate=\i as \x using -\t+2*\i] in {0,...,\steps}{
				\thetastateflipped[\x][0][1][#5]
			}
		}{
			\foreach \i[evaluate=\i as \x using -\t+2*\i] in {0,...,\steps}{      
				\thetastateflipped[-\x][0][1][#5]
			}
		}
	\end{scope}
}

\newcommandx{\eigenDL}[6][1=0,2=0,3=l,4=4,5=bertiniorange,6=topright]{
	\begin{scope}[shift={(#1,#2)}]
		\pgfmathsetmacro{\t}{#4}
		\ifthenelse{\equal{#3}{l}}{
			\eigenVL[0][0][l][\t][tr][init][#5][#6]
			\pgfmathsetmacro{\t}{#4-1}
			\rightriangle[1][0][\t][#5][#6]
			\drawinitstate[1][0][r][\t][#5]
		}{
			\begin{scope}[shift={(-0.5,0.5)}]
				\foreach \i[evaluate=\i as \x using \i, evaluate=\i as \y using \i] in {0,...,\t}{      
					\draw (\x,\y)--++(0.5,0);
					\draw[fill=white] (\x,\y) circle (0.15);
				}
			\end{scope}
		}
	\end{scope}
}

\newcommandx{\eigenDR}[6][1=0,2=0,3=l,4=4,5=bertiniorange,6=topright]{
	\begin{scope}[shift={(#1,#2)}]
		\pgfmathsetmacro{\t}{#4}
		\ifthenelse{\equal{#3}{r}}{
			\eigenVR[0][0][r][\t][tr][init][#5][#6]
			\pgfmathsetmacro{\t}{#4-1}
			\leftriangle[-1][0][\t][#5][#6]
			\drawinitstate[-1][0][l][\t][#5]
		}{
			\begin{scope}[shift={(0.5,0.5)}]
				\foreach \i[evaluate=\i as \x using \i, evaluate=\i as \y using \t-\i] in {0,...,\t}{      
					\draw (\x,\y)--++(0.5,0);
					\draw[fill=white] (\x+0.5,\y) circle (0.15);
				}
			\end{scope}
		}
	\end{scope}
}

\newcommandx{\idonpurity}[2][1=0,2=0]
{
	\begin{scope}[shift={(#1,#2)}]
		\draw[thick] (-0.5,0)--++(-0.1,0.1)--++(0,0.2)--++(0.1,-0.1);
		\draw[thick] (-0.5,0.4)--++(-0.1,0.1)--++(0,0.2)--++(0.1,-0.1);
		\draw[thick] (0.5,0)--++(0.1,0.1)--++(0,0.2)--++(-0.1,-0.1);
		\draw[thick] (0.5,0.4)--++(0.1,0.1)--++(0,0.2)--++(-0.1,-0.1);
	\end{scope}
}

\newcommandx{\swaponpurity}[2][1=0,2=0]
{
	\begin{scope}[shift={(#1,#2)}]
		\draw[thick] (-0.5,0)--++(-0.2,0.2)--++(0,0.6)--++(0.2,-0.2);
		\draw[thick] (-0.5,0.2)--++(-0.075,0.075)--++(0,0.2)--++(0.075,-0.075);
		\draw[thick] (+0.5,0)--++(+0.2,0.2)--++(0,0.6)--++(-0.2,-0.2);
		\draw[thick] (+0.5,0.2)--++(+0.075,0.075)--++(0,0.2)--++(-0.075,-0.075);
	\end{scope}
}

\newcommandx{\hook}[4][1=0,2=0,3=t,4=l]{
	\begin{scope}[shift={(#1,#2)}]
		\ifthenelse{\equal{#3}{t}}{
			\ifthenelse{\equal{#4}{l}}{\draw[thick] (0.5,-0.5) arc (45:-90:0.15);}{\draw[thick] (0.5,-0.5) arc (45:270:0.15);}
		}{\ifthenelse{\equal{#4}{l}}{\draw[ thick] (0.5,-0.5) arc (-45:90:0.15);}{\draw[ thick] (0.5,-0.5) arc (315:90:0.15);}
		}
	\end{scope}
}

\newcommandx{\hhook}[4][1=0,2=0,3=t,4=l]{
	\begin{scope}[shift={(#1,#2)}]
		\ifthenelse{\equal{#3}{t}}{
			\ifthenelse{\equal{#4}{l}}{\draw[thick] (0.5,-0.5) arc (-45:175:0.15);}{\draw[thick] (0.5,-0.5) arc (225:0:0.15);}
		}{\ifthenelse{\equal{#4}{l}}{\draw[ thick] (0.5,-0.5) arc (-45:180:-0.15);}{\draw[ thick] (0.5,-0.5) arc (45:-180:0.15);}
		}
	\end{scope}
}

\newcommandx{\Pproj}[3][3=$P_\Lambda$]{
\begin{scope}[shift={(#1-.5,#2-1)}]
\draw[thick,fill=white] (0,0)rectangle (1,2);
\draw[thick] (0,1.5)--(-.5,1.5);
\draw[thick] (1,1.5)--(1.5,1.5);
\draw[thick] (0,.5)--(-.5,.5);
\draw[thick] (1,.5)--(1.5,.5);
\node[scale=2] at (.5,1) {#3};
\end{scope}}

\definecolor{FcolU}{rgb}{0.71,0.78,0.91}
\definecolor{colLines}{rgb}{0.31,0.31,0.31}
\definecolor{colVMPSLines}{rgb}{0.11,0.11,0.11}
\definecolor{IcolUc}{rgb}{0.71,0.41,0.42}
\definecolor{IcolU}{rgb}{0.71,0.8,0.76}
\definecolor{IcolVMPSc}{rgb}{0.73,0.69,0.7}
\definecolor{IcolVMPS}{rgb}{0.81,0.77,0.78}
\definecolor{colObs}{rgb}{1.,1.,1.}

\newcommandx{\eightlegs}[2][1=0,2=0]{
	\begin{scope}[shift={(#1,#2)}]
		\foreach \x in {1,...,8}{
			\draw (\x, 0)--++(0,0.25);
			\draw[fill] (\x,0) circle (0.05);
		}
		\foreach \x in {1,3}{
			\pgfmathsetmacro\result{2*\x-1} 
			\node () at (\result,-0.5) {$i_{\x}$};
			\pgfmathsetmacro\result{2*\x}
			\node () at (\result,-0.5) {$j_{\x}$};	
		}
		\foreach \x in {2,4}{
			\pgfmathsetmacro\result{2*\x} 
			\node () at (\result,-0.5) {$i_{\x}$};
			\pgfmathsetmacro\result{2*\x-1}
			\node () at (\result,-0.5) {$j_{\x}$};	
		}
	\end{scope}
}

\newcommandx{\MPSinitialstate}[5][1=0,2=0,3=bertiniorange,4=topright,5=2]{
\begin{scope}[shift={(#1,#2)},rounded corners=1.5pt]
	\draw[black,thick,fill=#3] 
	(-0.25,.25)--++(.5,0)--++(0,-.3)--++(-.5,0)--cycle;
	\draw[thick] (-.25,.25)--++(-.25,.25);
	\draw[thick] (.25,.25)--++(.25,.25);
	\draw[very thick] (-1,.-.05)--++(2,0);
		\node at ({0},{0.085}) {\scalebox{1.}{{$#5$}}};
\end{scope}
}

\newcommandx{\Cmatrix}[6][1=0,2=0,3=2,4=bertiniorange,5=,6=2]{
	\pgfmathsetmacro\result{#3-1} 
	\begin{scope}[shift={(#1,#2)}]
		\foreach \i in {0,...,\result}
		{\foreach \j in {0,...,\i}
			{\roundgate[\i+\j][\i-\j][1][#6][#4][#6]}
		}
		\ifthenelse{\equal{#5}{init}}{
			\foreach \i in {0,...,#3}
			{
				\MPSinitialstate[-1+2*\i][-1][#4][][#6]
			}
		}{}
	\end{scope}
}

\newcommand{\mbcirc}[0]{
		\begin{tikzpicture}
			\draw[fill=black]  (0,0) circle  (0.12);
		\end{tikzpicture}}

\newcommand{\mcirc}{\mathbin{\scalerel*{\fullmoon}{G}}}

\newcommandx{\cstate}[4][1=0,2=0,3= ,4=white]{
	\begin{scope}[shift={(0,0)}]
				\draw[fill=#4,thick] (#1,#2) circle (0.15);
				\node[scale=1.1] at (#1,#2) {$#3$};
\end{scope}
}
\newcommandx{\sqrstate}[4][1=0,2=0,3= ,4=white]{
	\begin{scope}[shift={(#1,#2)}]
		\draw[thick,fill=#4] (-0.15,-0.15) rectangle (0.15,0.15) ;
		\node[scale=1.1] at (0,0) {$#3$};
	\end{scope}
}
\newcommand{\be}{\begin{equation}}
	\newcommand{\ee}{\end{equation}}
\pgfplotsset{
	colormap={bright}{rgb255=(251,51,255) rgb255=(255,128,0) rgb255=(0,0,255)
		rgb255=(255,0,0) rgb255=(128,255,0) rgb255=(204,204,0) rgb255=(127,0,255)
		rgb255=(0,0,0)}
}
\pgfplotsset{
	colormap={lightbluepalette}{ rgb255=(200, 240, 250) rgb255=(135, 206, 250) rgb255=(115, 194, 251)
		rgb255=(124,158,217) rgb255=(96, 130, 182)}
}

\theoremstyle{plain}        
\newtheorem{property}{Property}

\definecolorseries{foo}{hsb}{step}[hsb]{0,1,1}[hsb]{.618,0,0}

\newcommand{\SoPA}{School of Physics and Astronomy, University of Nottingham, Nottingham, NG7 2RD, UK}

\newcommand{\CQNE}{Centre for the Mathematics and Theoretical Physics of Quantum Non-Equilibrium Systems, University of Nottingham, Nottingham, NG7 2RD, UK}

\hypersetup{
	pdftitle={Quantum information spreading in generalised dual-unitary circuits},
	pdfsubject={Many-body quantum physics},
	pdfauthor={Alessandro Foligno, Pavel Kos, and Bruno Bertini},
	pdfkeywords={Many-body quantum physics, Entanglement, OTOC, Information spreading}
}
\begin{document}
	
	\title{Quantum information spreading in generalised dual-unitary circuits}
	
	\author{Alessandro Foligno}
	\affiliation{\SoPA} \affiliation{\CQNE}
	
	\author{Pavel Kos}
	\affiliation{Max-Planck-Institut f\"ur Quantenoptik, Hans-Kopfermann-Str. 1, 85748 Garching}
	
	\author{Bruno Bertini}
	\affiliation{\SoPA} \affiliation{\CQNE}
	
	% \date{\today}
	
	\begin{abstract}
		We study the spreading of quantum information in a recently introduced family of brickwork quantum circuits that generalises the dual-unitary class. These circuits are unitary in time, while their spatial dynamics is unitary only in a restricted subspace. First, we show that local operators spread at the speed of light as in dual-unitary circuits, i.e., the butterfly velocity takes the maximal value allowed by the geometry of the circuit. Then, we prove that the entanglement spreading can still be characterised exactly for a family of compatible initial states (in fact, for  an extension of the compatible family of dual-unitary circuits) and that the asymptotic entanglement slope is again independent on the R\'enyi index. Remarkably, however, we find that the entanglement velocity is generically smaller than one. We use these properties to find a closed-form expression for the entanglement-membrane line tension. 
	\end{abstract}
	
	\maketitle
	
	In recent years, quantum circuits have emerged as useful effective models to understand generic quantum many-body dynamics~\cite{hosur2016chaos, nahum2017quantum, bertini2018exact, nahum2018operator,chan2018solution,vonKeyserlingk2018operator,bertini2019entanglement,bertini2019exact,friedman2019spectral,li2019measurement,skinner2019measurement,rakovszky2019sub,zabalo2020critical}, and as concrete platforms for quantum simulation~\cite{brydges_probing_2019,elben_many-body_2019,elben_renyi_2018,pichler_measurement_2016,vermersch_probing_2019,vermersch_unitary_2018, aaronson_shadow_2018,huang_predicting_2020,ohliger_efficient_2013, keenan2022evidence, morvan2022formation}. From a theoretical point of view their most appealing feature is that, contrary to generic many-body systems in continuous time, their dynamics are sometimes amenable to analytical descriptions. This is particularly significant in light of the current lack of computational approaches able to efficiently characterise out-of-equilibrium quantum matter. 
	
	The approaches to obtain analytical insights in quantum circuits can be divided in two groups. The first involves introducing a certain degree of randomness in the system to simplify treatment~\cite{nahum2017quantum, chan2018solution, fisher2022random}. This approach is inspired by random matrix theory~\cite{mehta1991random} and has its most representative example in Haar-random circuits~\cite{nahum2017quantum}, which led to several important results on operator dynamics and information spreading~\cite{nahum2017quantum, vonKeyserlingk2018operator,  chan2018solution, khemani2018operator, rakovszky2018diffusive, zhou2020entanglement, wang2019barrier, fisher2022random, nahum2018operator,skinner2019measurement,li2019measurement,chan_unitary-projective_2019, friedman2019spectral, bertini2018exact,chan2018spectral, garratt2021manybody, garratt2021local, jonay2018coarsegrained, zhou2019emergent}. The second route, instead, is to derive exact results for special classes of systems obtained by imposing certain conditions on the elementary quantum gates without affecting the nature of the dynamics~\cite{bertini2019exact, klobas2021exact, prosen2021many, kos2021correlations, bertini2022exact, kos2023circuitsofspacetime, bertini2023exact}. The appeal of this second approach is that it is arguably more fundamental --- it allows one to study truly closed quantum many-body systems --- and its most representative example is that of dual-unitary (DU) circuits~\cite{bertini2019exact}. Importantly, the latter are not artificial theoretical abstractions: they can implement standard Floquet dynamics, e.g.\ the kicked Ising model~\cite{bertini2019entanglement, gopalakrishnan2019unitary}, and have been implemented in real-world quantum computers~\cite{chertkov2022holographic,mi_information_2021}.

	The defining property of DU circuits is that their local gates remain unitary upon exchanging the roles of space and time. This gives access to measures of quantum information spreading and operator growth that are notoriously hard to compute in general, see e.g., Refs.~\cite{claeys2020maximum, bertini2020scrambling,kos2023scrambling,rampp2023from, bertini2019entanglement, gopalakrishnan2019unitary, piroli2020exact, foligno2022growth, giudice2021temporal, foligno2023temporal, bertini2020operator, bertini2020operator2, reid2021entanglement, kos2023scrambling, ho2022exact, claeys2022emergentquantum, ippoliti2023dynamical, bertini2020scrambling}. %(examples include out-of-time correlators (OTOCs)~\cite{claeys2020maximum, bertini2020scrambling,kos2023scrambling,rampp2023from}, entanglement growth after quantum quenches~\cite{bertini2019entanglement, gopalakrishnan2019unitary, piroli2020exact, foligno2022growth}, temporal entanglement~\cite{giudice2021temporal, foligno2023temporal}, operator entanglement~\cite{bertini2020operator, bertini2020operator2, reid2021entanglement, kos2023scrambling}, deep thermalization~\cite{ho2022exact, claeys2022emergentquantum, ippoliti2023dynamical}, and tripartite information~\cite{bertini2020scrambling}). 
	Despite their solvability, DU circuits are provably quantum chaotic~\cite{bertini2018exact, bertini2021random} and display almost generic dynamics. The only macroscopic effect of dual unitarity is that it enforces maximal velocity for the spreading of quantum correlations. Specifically, in DU circuits both the velocity characterising operator spreading and the entanglement velocity of any state are equal to the speed of light~\cite{claeys2020maximum, piroli2020exact, foligno2022growth}. In fact, the second property implies conversely the dual unitarity~\cite{zhou2022maximal}.

	The fact that both scrambling and thermalisation occur at the fastest possible rate in DU circuits leaves a distinct mark on the dynamics of quantum information. This is true even at the coarse-grained level where quantum information spreading is described by the so-called entanglement membrane~\cite{jonay2018coarsegrained, zhou2019emergent, zhou2020entanglement} (see also~\cite{mezei2020exploring, agon2021bit}). DU circuits have been shown to exhibit an extremal, constant membrane line tension~\cite{zhou2020entanglement}, rather than the general convex function observed numerically in generic systems~\cite{zhou2020entanglement}, where scrambling and thermalisation occur at different, sub-maximal rates with the only constraint that the former occurs before the latter. The natural question is then whether the dual unitarity condition can somehow be weakened, leading to a more generic, yet solvable, quantum information flow.  
	
	In this Letter we address this question by characterising the dynamics of quantum information in a class of ``hierarchical generalisations'' of DU circuits recently proposed in Ref.~\cite{yu2023hierarchical}. 
	%The idea is to construct a hierarchy of multi-gate conditions, where going up in the hierarchy corresponds to weakening the constraint imposed on a single gate, with dual unitarity being at the bottom as the strongest one. 
	The idea is to construct a hierarchy of increasingly weaker conditions, with dual unitarity being at the bottom as the strongest one. Here we consider the second level of the hierarchy, DU2 from now on, and find the following results. First, the operator-scrambling velocity continues to be equal to one (in fact we show that this is true for all levels of the hierarchy). Second, the entanglement velocity is still independent on the R\'enyi index and can still be computed exactly. It is, however, generically \emph{sub-maximal}; {we interpret this by noting that the dual dynamics of the gate is proportional to an isometry. This constrains the exchange of correlations between different parts of the system and, ultimately, reduces the entanglement growth.} Finally, we recover these results by computing the entanglement membrane of DU2 and finding that it has non-trivial line tension. To the best of our knowledge the one provided here is the first explicit expression of a non-constant line tension derived in a clean, interacting system.

	\begin{figure*}
		\includegraphics[width=0.9\textwidth]{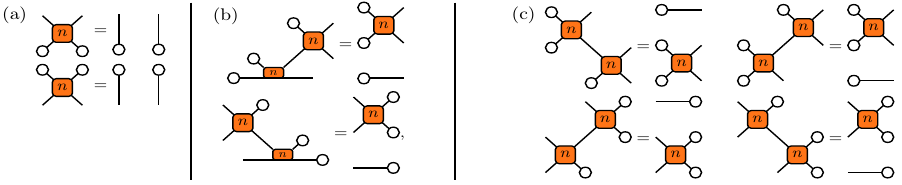}
		\caption{Graphical representations of unitarity (Left), compatibility relations for the initial state (cf.\ Eq.~\ref{eq:compatibilitycondition}) (Center), and DU2 conditions (cf.\ Eq.~\eqref{eq:2ndH}) (Right).}		
		\label{fig:conditions}
	\end{figure*}
	
	More specifically, we consider a one-dimensional quantum circuit, made of $2L$ sites of qudits (quantum systems with $d$ internal states), with a discrete, local unitary evolution.
	Neighbouring sites are by definition at distance of $1/2$ apart, and their positions, labelled by half-integers, take periodic values on a ring of length $L$ ($0\equiv L$). A single time step is determined by the unitary operator $\mathbb{U}=\mathbb{U}_e\mathbb{U}_o$ with 
	\begin{align}
		&\mathbb{U}_e=\bigotimes_{x\in \mathds{Z}_L} U_{x,x+1/2},& &\mathbb{U}_o=\bigotimes_{x\in \mathds{Z}_L} U_{x-1/2,x}\,,
		\label{eq:U}
	\end{align}
	$U_{a,b}$ is a two site unitary gate which acts on sites $a,b$. 
	
	We consider a subclass of unitary gates fulfilling the DU2 conditions~\cite{yu2023hierarchical}. These can expressed by defining a spacetime swapped gate $\tilde{U}$ as $\bra{kl}\tilde{U}\ket{ij}=\bra{lj}{U}\ket{ki}$. In terms of these gates the DU2 condition becomes (cf.\ Fig.~\ref{fig:conditions})
	\be
	\begin{aligned}
		\left(\tilde{U}\otimes \mathds{1}_d\right)\left(\mathds{1}_d\otimes\tilde{U}\tilde{U}^{\dagger} \right)\left(\tilde{U}^\dagger\otimes \mathds{1}_d\right)=	\tilde{U}\tilde{U}^\dagger \otimes \mathds{1}_d,\\
		\left(\mathds{1}_d\otimes \tilde{U}\right)\left(\tilde{U}\tilde{U}^{\dagger} \otimes\mathds{1}_d\right)\left(\mathds{1}_d\otimes \tilde{U}^\dagger\right)=	\mathds{1}_d	\otimes \tilde{U}\tilde{U}^\dagger ,
	\end{aligned}
	\label{eq:2ndH}
	\ee
	where $\mathds{1}_x$ is the identity on a space of dimension $x$. This property is satisfied for DU gates where $\tilde{U}$ is unitary, however, it admits also families of non-DU solutions~\cite{yu2023hierarchical}. Note that Eq.~\eqref{eq:2ndH} implies the validity of the analogous relations with $\tilde U^\dag$ and $\tilde U$ exchanged~\cite{yu2023hierarchical}.

	Let us begin computing the speed of operator spreading, i.e.\ the \emph{butterfly velocity} $v_B$. The latter can be quantified by looking at the following OTOC
	\begin{align}
		\!\!O_{\alpha\beta}(x,t)=1\!-\!\frac{1}{d^{2L}}\tr[\sigma_0^{(\alpha)}(t)\sigma_x^{(\beta)}(0)\sigma_0^{(\alpha)}(t)\sigma_x^{(\beta)}(0)],
		\label{eq:OTOCdef}\end{align}
	where $\{\sigma^{(\alpha)}\}_{\alpha=1,\ldots d^2-1}$ are a basis for local traceless hermitian operators~\footnote{We take them to be orthogonal with respect to the Hilbert-Schmidt product, i.e., $\tr\smash{[\sigma^{(\alpha)}\sigma^{(\beta)}]}=d \delta_{\alpha,\beta}$.}. In chaotic systems, this OTOC approaches asymptotically one for ${\abs{x}\le v_B t}$, and $0$ otherwise. In particular, Haar random circuits have $v_B={(d^2-1)}/{(d^2+1)}$~\cite{nahum2018operator,vonKeyserlingk2018operator}, while DU circuits ${v_B=1}$~\cite{claeys2020maximum, bertini2020scrambling}. Note that the latter is the largest possible $v_B$ because the strict causality encoded in Eq.~\eqref{eq:U} assures $O_{\alpha\beta}(x,t)=0$ for $|x|>t$.

	To compute $v_B$ we use the strategy of Refs.~\cite{claeys2020maximum, bertini2020scrambling}. Namely, we compute the limit $x,t\to\infty$ with $x_-=t-x$ fixed: if $O_{\alpha\beta}(x,t)$ is non-zero in this limit we have $v_B=1$ otherwise $v_B<1$~\footnote{Here we assume parity symmetry. If there is no parity symmetry one has to consider also the limit $|x|,t\to\infty$ with $x_+=t+x$ fixed. The treatment is completely analogous.}. The limit is conveniently computed writing $1-O_{\alpha\beta}(x,t)$ in terms of a suitable transfer matrix and expressing its asymptotic scaling in terms of the transfer-matrix fixed points. This procedure becomes particularly transparent by introducing a diagrammatic representation, similar to the one used in tensor networks, where one depicts single local gates as boxes with legs corresponding to the qudits they act on, see e.g.~\cite{gopalakrishnan2019unitary, bertini2019exact, bertini2020scrambling}. In particular, since here we are interested in multi replica quantities we consider a graphical representation of ``folded'' quantum gates, i.e., tensor products of $n$ replicas of $U$ and its conjugate
	\begin{align}
		\label{eq:localgate}
		\fineq{\roundgate[0][0][.8][topright][bertiniorange][n]} \equiv \left(U\otimes U^*\right)^{\otimes n}\,.
	\end{align}
	The Hilbert spaces associated to each leg have dimension $d^{n}$. For $|x|<t$ one can express the quantity of interest in terms of \eqref{eq:localgate} as follows \footnote{In Eq.~\eqref{eq:diagrammaticOTOC} we assumed $x$ to be half-integer; the case of integer $x$ leads to identical considerations}
	\be
	\!\!\!\!\!\!\!1\!-\!O_{\alpha\beta}(x,t)\!=\!\frac{1}{d^{2t}}\fineq[1.4ex][0.55][1]
	{	\begin{scope}[rotate around={-45:(0,0)}]
			\foreach \i in {0,...,4}
			{
				\foreach \j in {0,...,2}
				{
					\roundgate[\i+\j][\i-\j][1][topright][bertiniorange][2]
					\cstate[\i-0.5][\i+0.5][]
					\sqrstate[3+\i-0.5][-2+\i-0.5]
				}
			}
			\foreach \i in {0,...,4}
			{
				\foreach \j in {0,...,1}
				{
					\sqrstate[\j-0.5][-\j-0.5]
					\cstate[4+2-\j+0.5][2+\j+0.5]
				}	
			}
			\draw[fill=black,shift={(1.5,-2.5)}] (-.1,-.1) rectangle (.1,.1);
			\draw[fill=black,shift={(4.5,4.5)}] (0,0) circle (.12);
		\end{scope}
		\node[scale=1.2] at	(-.75,-3.25) {${\beta}$};
		\node[scale=1.2] at	(6.4,0.3) {${\alpha}$};
		\draw [decorate, decoration = {brace,,mirror}]   (0,-4)--(5.75,-4) ;
		\draw [decorate, decoration = {brace}]   (6.6,0.1)--(6.6,-2.9) ;
		\node[scale=1.2] at (3.,-4.5) {$x_+$};
		\node[scale=1.2] at (7.1,-1.5) {$x_-$};}\!.
	\label{eq:diagrammaticOTOC}
	\ee	
	Here $x_\pm=t\pm x$, joined legs imply matrix product and we introduced a graphical representation for two different index contractions that can be seen as states in the replicated Hilbert space, i.e.
		\be
		\!\!\!\ket{\fineq[-0.8ex][0.7][1]{					\cstate[0][0][]}}=\sum_{i_k=1}^d\ket{i_1i_1\dots i_n i_n},\quad \ket{\fineq[-0.8ex][0.7][1]{\sqrstate[0][0][]}}=\sum_{i_k=1}^d\ket{i_n i_1 i_1 \ldots i_n},\label{eq:staggeredstates}
		\ee
		where we used the shorthand notation $\ket{i_1i_2\ldots i_{2n}}\equiv \ket{i_1}\otimes\ket{i_2}\otimes\cdots\otimes\ket{i_{2n}}$. These states are neither orthogonal nor normalised and one has $\braket{\fineq[-0.8ex][0.7][1]{\cstate[0][0][]}}{\fineq[-0.8ex][0.7][1]{\sqrstate[0][0][]}}=d$
		 and $\braket{\fineq[-0.8ex][0.7][1]{\cstate[0][0][]}}=\braket{\fineq[-0.8ex][0.7][1]{\sqrstate[0][0][]}}=d^n$. 
		Similarly we set 
		\be
		\ket{\mbcirc_{\alpha}}=(\sigma^{(\alpha)} \otimes\mathds{1}_d)^{\otimes n} \ket{\fineq[-0.8ex][0.7][1]{\cstate[0][0][]}},\,\,
		\ket{\fineq[-0.8ex][0.9]{\draw[fill=black] (-.1,-.1) rectangle (.1,.1);}_\beta}=(\sigma^{(\beta)} \otimes\mathds{1}_d)^{\otimes n} \ket{\fineq[-0.8ex][0.7][1]{\sqrstate[0][0][]}}.\label{eq:boundarystates}
		\ee
		For the sake of compactness, we suppressed the $n$-dependence from the l.h.s.\ of Eqs.~\eqref{eq:staggeredstates} and \eqref{eq:boundarystates} because in the diagrams the value of $n$ is specified in the gates.
	
	From the diagram \eqref{eq:diagrammaticOTOC} we see that $1-O_{\alpha\beta}(x,t)$ is written as 
	\begin{align}
		1-O_{\alpha\beta}(x,t)=\frac{1}{d^{x_-}}	\mel{\smash{\fineq[-0.8ex][0.7][1]{\sqrstate[0][0][]}\cdots \fineq[-0.8ex][1]{\draw[fill=black] (-.1,-.1) rectangle (.1,.1);}_\beta}}{\smash{T_{x_-}^{x_+}}}{\smash{\mbcirc_\alpha \cdots \fineq[-0.8ex][0.7][1]{\cstate[0][0][]}}},\label{eq:expval}
	\end{align} where we introduced 
	\begin{align}
		T_x=\frac{1}{d}\,\,\fineq[-3ex][0.65][1]{	
			\begin{scope}[rotate around={45:(0,0)}]
				\foreach \i in {0,...,3}
				{
					\roundgate[-\i][\i][1][topright][bertiniorange][2]
				}
				\sqrstate[.5][-.5]
			\end{scope}
			\cstate[-5][0]
			\draw [decorate, decoration = {brace}]   (-4.5,1)--++(4.8,0) ;
			\node[scale=1.5] at (-2.05,1.4) {$x$};}
	\end{align}
	This matrix has maximal eigenvalue $1$ fixed by unitarity~\cite{claeys2020maximum, bertini2020scrambling} . Therefore, in the limit of interest we can replace it by $\sum_i \ketbra{r_i}{l_i}/\braket{l_i}{r_i}$ in Eq.~\eqref{eq:expval}, where $\bra{l_i}$ and $\ket{r_i}$ denote respectively its right and left fixed points. Generically $T_{x_-}$ has a unique left and right fixed points guaranteed by unitarity, i.e., $\ket{\fineq[-0.8ex][0.7][1]{\cstate[0][0][]}\ldots\fineq[-0.8ex][0.7][1]{\cstate[0][0][]}}, \bra{\fineq[-0.8ex][0.7][1]{\sqrstate[0][0][]}\ldots\fineq[-0.8ex][0.7][1]{\sqrstate[0][0][]}}$ 
	(this can be seen graphically using the conditions in Fig.~\ref{fig:conditions}a). The latter, however, contribute with a one in the r.h.s. of Eq.~\eqref{eq:expval}. Therefore, the OTOC can have a non-zero value in the limit of interest only if $T_x$ has at least another nontrivial fixed point for some $x_-$.

	We now show that DU2 provides such an additional fixed point. We consider $x_-=1$ and employ Eq.~\eqref{eq:2ndH} (cf.\ Fig.~\ref{fig:conditions}) to show that
	\begin{align}
		\bra{l}=\bra{\fineq[-0.8ex][0.7][1]{\cstate[0][0][]}} T_1=
		\fineq[-.2ex][0.65][1]
		{
			\begin{scope}[rotate around={45:(0,0)}]
				\foreach \i in {0}
				{
					\roundgate[-\i][\i][1][topright][bertiniorange][2]
					\cstate[-0.5][0.5]
					\cstate[-0.5][-0.5]
					\sqrstate[.5][-0.5]
				}
			\end{scope}
		}\,,\,\,\, \ket{r}=T_1\ket{\fineq[-0.8ex][0.7][1]{\sqrstate[0][0][]}} =
		\fineq[-1ex][0.65][1]
		{
			\begin{scope}[rotate around={45:(0,0)}]
				\foreach \i in {0}
				{
					\roundgate[-\i][\i][1][topright][bertiniorange][2]
					\cstate[-0.5][0.5]
					\sqrstate[0.5][-0.5]
					\sqrstate[0.5][0.5]
				}
			\end{scope}
		},\label{eq:eigenvectors}
	\end{align}
	are fixed points of $T_1$. These vectors have non-zero overlap with states orthogonal to $\ket{\fineq[-0.8ex][0.7][1]{\sqrstate[0][0][]}}$ 
	and $\ket{\fineq[-0.8ex][0.7][1]{\cstate[0][0][]}}$ 
	respectively, and therefore contribute to the r.h.s.\ of Eq.~\eqref{eq:expval} for some $\alpha$ and $\beta$. In particular, defining 
	\begin{align}
		\ket{\mbcirc}\equiv d\ket{\fineq[-0.8ex][0.7][1]{\sqrstate[0][0][]}}-\ket{\fineq[-0.8ex][0.7][1]{\cstate[0][0][]}},\qquad\ket{\fineq[-0.8ex][0.9]{\draw[fill=black] (-.1,-.1) rectangle (.1,.1);}}=d\ket{\fineq[-0.8ex][0.7][1]{\cstate[0][0][]}}-\ket{\fineq[-0.8ex][0.7][1]{\sqrstate[0][0][]}}\label{eq:mbcircdef}
	\end{align} 
	we have~\footnote[20]{See the Supplemental Material for: (i) an explicit calculation of the scalar product in Eq.~\eqref{eq:scalarproduct}; (ii) A characterisation of R\'enyi entropies at early times, i.e., for $L\geq L_A+2t \geq 4t$ and a proof of Eq.~\eqref{eq:factorizedrenyis}; (iii) A proof that the matrix $\tilde{U}\tilde{U}^\dag$ has flat spectrum; (iv) The parameterisation considered in our numerical experiments.}
	\begin{align}
		\braket{l}{\fineq[-0.8ex][0.9]{\draw[fill=black] (-.1,-.1) rectangle (.1,.1);}}=\braket{\mbcirc}{r}=d^4\left(1-\frac{1}{n_\Lambda}\right),
		\label{eq:scalarproduct}
	\end{align}
	where $n_\Lambda=1,\ldots, d^2$ depending on the specific choice of the DU2 gate. The case $n_\Lambda=1$ corresponds to a trivial gate with no entangling power, i.e., $U=u \otimes v$. Assuming $n_\Lambda \ne 1$, we find that $\lim_{t\to\infty}O_{\alpha\beta}(t-1,t)\neq 0$ and the butterfly velocity for DU2 gates is indeed maximal. This reasoning can be extended to the full hierarchy of Ref.~\cite{yu2023hierarchical} (focussing again on entangling  gates) by considering 
		\begin{align}  
			\bra{l}=\bra{\fineq[-0.8ex][0.7][1]{\cstate[0][0][]}}\left(T_1\right)^{m-1},\quad \ket{r}=\left(T_1\right)^{m-1}\ket{\fineq[-0.8ex][0.7][1]{\sqrstate[0][0][]}}, 
		\end{align}
		for gates at the $m$-th level of the hierarchy. This is the first main result of this letter: all generalised dual-unitary circuits have butterfly velocity pinned at one. This property shows that generalised dual-unitary circuits can never be fully generic when it comes to operator spreading. A maximal butterfly velocity, however, is not a very constraining feature by itself. Intuitively one can always think of achieving it by applying enough steps of tensor network renormalisation to the quantum circuit~\cite{evenbly2015tensor}.

	Let us now move on to the calculation of the entanglement velocity by looking at a quantum quench from a class of initial states compatible with the DU2 property~\cite{yu2023hierarchical}. The latter are expressed as two-site shift invariant matrix product states (MPS) with bond dimension $\chi$, namely 
	\begin{align}
		\!\!\ket{\Psi_0}\!=\!
		\frac{1}{d^{L/2}}\!\!\sum_{i_k=1}^d \!\tr\!\smash{[M^{[i_1 i_2]}\ldots M^{[i_{2L-1} i_{2L}]}]} \ket{i_1\ldots i_{2L}}\!,
	\end{align}
	where the $\chi\times\chi$ matrices $\{M^{[a,b]}\}$ are chosen such that the MPS transfer matrix $\sum_{i,j=1}^d M^{[i,j]} \otimes (M^{[i,j]})^*$ has maximal, non-degenerate eigenvalue $d$, corresponding to eigenvectors $\bra{\Omega_L}$ and $\ket{\Omega_R}$. Moreover, the $d \chi\times d\chi$ matrices $W$ with elements $\mel{ia}{W}{jb}=\mel{i}{M^{(ab)}}{j}$ fulfil 
	\be
	\begin{aligned}
		\left(\tilde{U}\otimes \mathds{1}_\chi\right)\left(\mathds{1}_d \otimes W W^{\dagger} \right)\left(\tilde{U}^\dagger\otimes \mathds{1}_\chi\right)=	\tilde{U}\tilde{U}^\dagger \otimes \mathds{1}_{\chi},\\
		\left(\tilde{U}{^\dagger}\otimes \mathds{1}_\chi\right)\left(\mathds{1}_d\otimes W^\dagger W \right)\left(\tilde{U}\otimes \mathds{1}_\chi\right)=	\tilde{U}^\dagger\tilde{U} \otimes \mathds{1}_{\chi},
	\end{aligned} 
	\label{eq:compatibilitycondition}	
	\ee
	see Fig.~\ref{fig:conditions}b. These relations are solved by the initial states compatible with dual unitarity introduced in Ref.~\cite{piroli2020exact}, however, these are not the only solutions. Note that \eqref{eq:compatibilitycondition} also implies $\ket{\Omega_R}=(\bra{\Omega_L})^\dag=\sum_{i=1}^\chi \ket{i i}$. 
	
	Considering the evolution of the R\'enyi entropies of a block $A$ (of length $L_A$), namely  
	\begin{align}
		S^{(n)}_A(t)=\frac{1}{1-n}\log(\tr_A[\rho_A^n(t)]),
		\label{eq:renyientropy}
	\end{align}
	where $\rho_A(t)$
	 is the state at time $t$ reduced to $A$, and taking the limit $L\to\infty$ followed by $L_A\to\infty$ we obtain~\cite{Note20} 
	\begin{align}
		&\lim_{L_A\to\infty}\lim_{L\to\infty}S^{(n)}_A(t) = \frac{2 n}{n-1}\log {\chi d^{(2t+1)}} \notag \\
		&-\frac{2}{n-1}\log\!\!\!\!\! 		\fineq[-0.8ex][0.55][1]{
			\Cmatrix[0][0][4][bertiniorange][init][n]
			\foreach \i in {0,...,4}
			{
				\cstate[-1.5+\i][-.5+\i]
			}
			\cstate[-2][-.5-.55]
			\foreach \i in {0,...,4}
			{
				\sqrstate[-1.5+10-1-\i][-.5+\i][]
			}
			\sqrstate[-2+10][-.5-.55][]
		}\,,
		\label{eq:factorizedrenyis}
	\end{align}
	where we introduced the following symbol for the tensor product of $n$ copies of $W$ and $W^*$
	\begin{align}
		\fineq[-0.4ex][0.8][1]{				\MPSinitialstate[0][0][bertiniorange][topright][n]}=\left(W\otimes W^*
		\right)^{\otimes n},
	\end{align}
	and the thicker line at the bottom corresponds to $2n$ copies of the MPS' auxiliary space.
	
	The physical meaning of \eqref{eq:factorizedrenyis} is that for early times ($4t\leq L_A+2t \leq L$) the entanglement between $A$ and the rest is only produced at the two boundaries between the two subsystems and the latter are causally disconnected~\cite{bertini2022entanglement}.

	Next we observe that, by repeated applications of the DU2 property and the diagrammatic version of the compatibility condition \eqref{eq:compatibilitycondition} (Fig.~\ref{fig:conditions}b and c) we can simplify the triangular diagram in Eq.~\eqref{eq:factorizedrenyis} to
		\begin{align}
			\left(\fineq[-0.8ex][0.65][1]{
				\foreach \i in {0}
				{
					\roundgate[0][0+2*\i][1][topright][bertiniorange][n]
					\cstate[-.5][0.5+2*\i][]
					\cstate[-.5][-0.5+2*\i][]
					\sqrstate[.5][0.5+2*\i]
					\sqrstate[.5][-0.5+2*\i]
				}
			}\right)^t \times
			\fineq[-0.8ex][0.65][1]{
				\MPSinitialstate[0][-2][bertiniorange][topright][n]
				\cstate[-.5][0.5-2]
				\cstate[-1][-0.05-2]
				\sqrstate[.5][0.5-2]
				\sqrstate[1][-0.05-2]	
			}\,.	\label{eq:entanglementsimplification}
	\end{align}
	This gives the following asymptotic entanglement velocity  
	\be
	\!\!\!\!v^{\!(n)}_E\!\!\equiv\!\! \lim_{t\rightarrow\infty}\lim_{L_A\to\infty}\lim_{L\to\infty}\!\frac{S^{(n)}_A(t)}{4 t \log(d) }\!\!=\!\frac{\log\tr[\smash{(\widetilde{U}\widetilde{U}^\dagger/d^2)^n}]}{2(1-n)\log(d)}
	\label{eq:entanglementspeedgeneral}.
	\ee
	This result generalises the one for DU circuits~\cite{bertini2019entanglement, piroli2020exact}, i.e.\ $v^{(n)}_E\!=\!1$, which is recovered setting $\tilde{U}\tilde{U}^\dagger= \mathds{1}_{d^2}$. Remarkably, however, Eq.~\eqref{eq:entanglementspeedgeneral} continues to be $n$ independent for all DU2 circuits. Indeed the spectrum of the matrix $\tilde{U}\tilde{U}^\dagger$ is \emph{constant} for all DU2 gates~\cite{Note20}: 	
	\begin{property}
		\label{prop:p1}
		For DU2 circuits the eigenvalues of $\tilde{U}\tilde{U}^\dagger$ are all either equal to $0$ or to a positive constant $\Lambda^2$. 
	\end{property}
	
	\noindent Since the trace of $\tilde{U}\tilde{U}^\dagger$ is fixed by the unitarity of $U$, the dimension of the non-trivial eigenspace, $n_\Lambda = 1, \ldots, d^2$, is such that $n_\Lambda \Lambda^2=d^2$. This allows us to rewrite Eq.~\eqref{eq:entanglementspeedgeneral} as
	\begin{align}
		v_E^{(n)}=\frac{\log(\frac{d}{\Lambda})}{\log(d)}= \frac{\log(n_\Lambda)}{2\log(d)}, \qquad n_\Lambda = 1, \ldots, d^2.
		\label{eq:finalformula}
	\end{align}
	This exact expression represents our second main result and, in contrast with that on the butterfly velocity, cannot be directly extended to the full hierarchy of Ref.~\cite{yu2023hierarchical}: beyond DU2 the triangular diagram in Eq.~\eqref{eq:factorizedrenyis} does not simplify and Property~\ref{prop:p1} does not hold. In fact, the validity of Property~\ref{prop:p1} seems to be the key to solvability as it implies that the space-time swapped gate $\tilde{U}$ is proportional to an isometry. Consequently, when swapping space and time the dynamics are unitary in a reduced subspace. This reduction lowers the entanglement velocity, which can now attain different discrete values (but not arbitrary ones as in the generic case).

\begin{figure}[t]		
\includegraphics[scale=1]{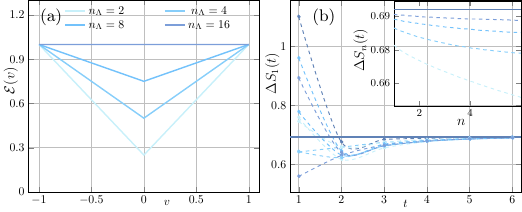}
\caption{(Left) Line tension of DU2 circuits [cf.\ Eq.~\eqref{eq:membranetension}], for $d=4$, and $n_\Lambda$ taking all possible values (except the trivial value $1$) corresponding to different gates. (Right) Entanglement slopes from random dimer-product states compared with the expected slope from solvable states (solid line). Inset: Slope of R\'enyi entropies for increasing values of $t$ (darker shade) as a function of the R\'enyi index $n$, compared to the $n-$independent result from solvable states (solid line).}	
\label{plot:membraneandgrowth}
\end{figure}

	We now recover our exact results using the entanglement membrane picture~\cite{jonay2018coarsegrained, zhou2019emergent, zhou2020entanglement}. The idea of this approach is to view the entanglement as the energy of a coarse-grained curve (which depends only on its slope). Namely, one can write a R\'enyi entropy as $S_n(x,t)=\min_y 	 \left(\mathcal{E}_n\left({(x-y)}/{t}\right)+ S_n(y,0)\right)$~\cite{jonay2018coarsegrained}. The function $\mathcal{E}_n\left(v\right)$ can be computed by evaluating the scaling limit of a suitable matrix element in the replicated space~\cite{zhou2020entanglement}. More precisely, we have 
	\begin{align}
		\!\!e^{\frac{t\mathcal{E}_n(v) \log d}{(1-n)^{-1}}} \!\! \simeq \! \frac{1}{d^{2nt}}\fineq[-0.4ex][0.55]{
			\begin{scope}[rotate around={-45:(0,0)}]
				\foreach \i in {0,...,2}
				\foreach \j in {0,...,4}
				{
					\roundgate[-\i-\j][\i-\j][1][topright][bertiniorange][n]
					\cstate[-\i+0.5][\i+0.5]
					\sqrstate[-\i-1.5-3][\i-1.5-3]
					\cstate[-\j+0.5][-\j-0.5]
					\sqrstate[-\j-2.5][3-\j-0.5]
				}
			\end{scope}
			\draw [decorate, decoration = {brace}, thick,shift={(0,0)}]   (1,3.1) --++ (0,-3.5);
			\draw [decorate, decoration = {brace}, thick,shift={(0,0)}]   (-6,3.8) --++ (6.3,0);
			\node[scale=1.3] at (-2.75,4.2) {$x_+$};
			\node[scale=1.3] at (1.5,1.3) {$x_-$};
		} ,\label{eq:tensiondef}
	\end{align}
	where $\simeq$ denotes equality at leading order in $t$. The matrix element on the r.h.s.\ is typically very hard to evaluate analytically and closed form expressions have only been found in the presence of randomness and for large $d$~\cite{jonay2018coarsegrained, zhou2019emergent, gong2022coarse}, in the dual-unitary case~\cite{zhou2020entanglement}, or for holographic quantum field theories~\cite{mezei2018membrane}. In our case, instead, the calculation is straightforward: Eq.~\eqref{eq:tensiondef}  can be explicitly contracted using the DU2 property (cf.\ Fig.~\ref{fig:conditions}c) starting from the top-left and bottom-right corners. A direct application of Property~\ref{prop:p1} then leads to our third main result
		\begin{align}
			\mathcal{E}_{n}(v)={}\left(\abs{v}+\frac{1-\abs{v}}{2}\frac{\log(n_\Lambda)}{\log(d)}\right).\label{eq:membranetension}
	\end{align}
	We see that $\mathcal{E}_{n}(v)$ shows a non-trivial dependence on $v$ but is convex as it should be for consistency. Specifically, it is generically linear in $|v|$ and becomes constant only in the DU case $n_\Lambda=d^2$ (see Fig.~\ref{plot:membraneandgrowth}a). As shown in Ref.~\cite{mezei2018membrane}, this form maximizes the growth of the entanglement for fixed $v_E$ and $v_B$. This extremal form of the membrane is again connected with the isometric nature of the space-time swapped dynamics.
	
	Eq.~\eqref{eq:membranetension} allows us to recover our exact results for butterfly and entanglement velocities within the membrane approach~\cite{jonay2018coarsegrained}: the solution to $\mathcal{E}(v)=v$ is indeed ${v=\!v_B\!=1}$ and $v^{(n)}_E\!=\!\mathcal{E}(v=0)$ coincides with Eq.~\eqref{eq:finalformula}. In fact, the membrane approach suggests that this result does not depend on the initial state, as long as it is low entangled, and should apply also for states that do not satisfy Eq.~\eqref{eq:compatibilitycondition}. This is confirmed by our numerics, see Fig.~\ref{plot:membraneandgrowth}b.

	In this Letter we presented an explicit characterisation of the quantum information flow in a class of unitary circuits, dubbed DU2 circuits, that generalises dual unitary circuits. We showed that although local operators spread at the maximal speed and the entanglement spectrum after a quench is asymptotically flat, the entanglement velocity is submaximal and its value depends on additional properties of the gate. Finally, we showed that these results can be recovered using the entanglement membrane approach by deriving an exact expression for the line tension.

	Our exact results put DU2 circuits forward as a general class of chaotic, yet solvable quantum circuits characterised by local and isometric space dynamics. Some immediate questions for future research are to establish whether DU2 circuits enjoy the same degree of solvability as dual-unitary circuits by investigating different properties (e.g.\ spectral correlations or temporal entanglement) and whether one can achieve some form of exact solvability by imposing even weaker constraints on the space dynamics, perhaps reducing its degree of locality.
 { Another interesting generalization would be to consider dynamics which are different isometric evolutions in the time and space directions. This could give possible new solvable families of circuits with measurements.}
	
	\begin{acknowledgments}
		
		We thank Katja Klobas for collaboration in the early stages of this project and for valuable comments on the manuscript. We also thank Xie-Hang Yu and Sarang Gopalakrishnan for useful discussions. We acknowledge financial support from the Royal Society through the University Research Fellowship No.\ 201101 (A.\ F.\ and B.\ B.). P.\ K.\ is supported by the Alexander von Humboldt Foundation. We warmly acknowledge the hospitality of the Simons Center for Geometry and Physics during the program ``Fluctuations, Entanglements, and Chaos: Exact Results'' where part of this work has been performed.
		
		{\it Note Added.} While this manuscript was being finalised, we became aware of the related work~\cite{rampp2023entanglement}, which will soon appear on arXiv.
		
	\end{acknowledgments}
	\bibliography{bibliography}

	%%%%%%%%%%%%%%%%%%%%%%%%%%%%%%%%%%%%%%%%%%%%
	\onecolumngrid
	\break
	
	\begin{center}
		{\large \bf Supplemental Material: \\
				Information spreading in generalised dual-unitary circuits}
	\end{center}
	\appendix
	
\onecolumngrid
Here we report some useful information complementing the main text. In particular

\begin{itemize}
  \item[-] In Sec.~\ref{sec:OTOC} we compute explicitly the scalar products $	\braket{l}{\fineq[-0.8ex][0.9]{\draw[fill=black] (-.1,-.1) rectangle (.1,.1);}}=\braket{\mbcirc}{r}$ in Eq.~\eqref{eq:scalarproduct} of the main text.
  \item[-] In Sec.~\ref{sec:entanglement} we characterise R\'enyi entropies at early times, i.e., for $L\geq L_A+2t \geq 4t$ and prove Eq.~\eqref{eq:factorizedrenyis} of the main text. 
    \item[-] In Sec.~\ref{sec:app1} we show that the matrix $\widetilde{U}\widetilde{U}^\dagger$ has flat spectrum.
    \item[-] In Sec.~\ref{sec:parameterisation} we report the parameterisation used for our numerical experiments. 
\end{itemize}

\section{Explicit calculation of the scalar product in Eq.~(\ref{eq:scalarproduct})}
\label{sec:OTOC}
Consider the vectors $\ket{l}$ and $\ket{\fineq[-0.8ex][0.9]{\draw[fill=black] (-.1,-.1) rectangle (.1,.1);}}$ in Eqs.~\eqref{eq:eigenvectors} and \eqref{eq:mbcircdef}. Their scalar product can be compute by decomposing $\ket{\fineq[-0.8ex][0.9]{\draw[fill=black] (-.1,-.1) rectangle (.1,.1);}}$
\begin{align}
	\braket{l}{\fineq[-0.8ex][0.9]{\draw[fill=black] (-.1,-.1) rectangle (.1,.1);}}=d\braket{l}{\fineq[-0.8ex][0.9]{\cstate[0][0]}}-\braket{l}{{\fineq[-0.8ex][0.9]{\sqrstate[0][0]}}}=d\,\fineq[-0.8ex][0.75][1]
		{
			\begin{scope}[rotate around={45:(0,0)}]
				\foreach \i in {0}
				{
					\roundgate[-\i][\i][1][topright][bertiniorange][2]
          			\cstate[-0.5][0.5]
     			\cstate[-0.5][-0.5]
        			\sqrstate[.5][-0.5]
     			\cstate[0.5][0.5]
				}
			\end{scope}
		}-\fineq[-0.8ex][0.75][1]
		{
			\begin{scope}[rotate around={45:(0,0)}]
				\foreach \i in {0}
				{
					\roundgate[-\i][\i][1][topright][bertiniorange][2]
          			\cstate[-0.5][0.5]
     			\cstate[-0.5][-0.5]
        			\sqrstate[.5][-0.5]
     			\sqrstate[0.5][0.5]
				}
			\end{scope}
		}.
\end{align}
The first term can be immediately simplified using unitarity and obtaining $d^4$, while the second one can be written in terms of $\tilde{U}$ as
\begin{align}
	\braket{l}{\mbcirc}=d^4-\tra{\left(\tilde{U}\tilde{U}^\dagger\right)^2}=d^4\left(1-\frac{1}{n_\Lambda}\right),
\end{align}
where we used Eq.~\eqref{eq:spectrumUtilde} to simplify the second term; identical considerations apply to $\braket{\fineq[-0.8ex][0.9]{\draw[fill=black] (-.1,-.1) rectangle (.1,.1);}}{r}$.

\section{Entanglement Dynamics at Early Times}
\label{sec:entanglement}

In order to compute the R\'enyi entropy from Eq.~\eqref{eq:renyientropy}, we can express the trace of the $n$-th power of the reduced density matrix as a contracted network:
\begin{align}
	&	\tra[]{\rho_A^n}	=\frac{1}{d^{nL}}\fineq[-0.8ex][0.55][1]{
		\foreach \i in {0,...,6}
		{
			\MPSinitialstate[\i*2][0][bertiniorange][topright][n]
			\ifnumless{\i}{2}{
				\foreach \j in {\i,...,5}
				{
					\roundgate[\j*2+3-\i-2][\i+1][1][topright][bertiniorange][n]
				}
			}
			{}
		}
		\foreach\i in {0,1,2}
		{	
			\cstate[-.5+\i][.5+\i]\cstate[12.5-\i][.5+\i]
		}
		\foreach\i in {0,...,7}
		{	
			\sqrstate[2.5+\i][2.5][]
		}
				\foreach \i in {1,...,2}
{
	\MPSinitialstate[-\i*2][0][bertiniorange][topright][n]
	\cstate[-.5-\i*2][.5]
	\cstate[.5-\i*2][.5]
	\MPSinitialstate[12+\i*2][0][bertiniorange][topright][n]
	\cstate[12+-.5+\i*2][.5]
	\cstate[12+.5+\i*2][.5]
}
\node[scale=1.5] at (17.5,-.1) {$\ldots$};
\node[scale=1.5] at (-5.5,-.1) {$\ldots$};
	}\notag\\&= 
\frac{1}{d^{nL}}	\fineq[-0.8ex][0.55][1]{
		\foreach \i in {0,...,6}
		{
			\MPSinitialstate[\i*2][0][bertiniorange][topright][n]
			\ifnumless{\i}{2}{
				\foreach \j in {\i,...,1}
				{
					\roundgate[\j*2+3-\i-2][\i+1][1][topright][bertiniorange][n]
					\roundgate[\j*2+3-\i+6][\i+1][1][topright][bertiniorange][n]
				}
			}
			{}
		}
		\foreach\i in {0,1,2}
		{	
			\cstate[-.5+\i][.5+\i]\cstate[12.5-\i][.5+\i]
		}
		\foreach\i in {0,...,2}
		{	
			\sqrstate[2.5+\i][2.5-\i]
			\sqrstate[9.5-\i][2.5-\i][]
		}
				\foreach \i in {1,...,2}
		{
			\MPSinitialstate[-\i*2][0][bertiniorange][topright][n]
			\cstate[-.5-\i*2][.5]
			\cstate[.5-\i*2][.5]
			\MPSinitialstate[12+\i*2][0][bertiniorange][topright][n]
		\cstate[12+-.5+\i*2][.5]
		\cstate[12+.5+\i*2][.5]
		}
		\node[scale=1.5] at (17.5,-.1) {$\ldots$};
		\node[scale=1.5] at (-5.5,-.1) {$\ldots$};
		\sqrstate[5.5][.5]
		\sqrstate[6.5][.5]
		\draw [decorate, decoration = {brace}]   (5,-.5)--(-1,-.5) ;
		\draw [decorate, decoration = {brace},shift={(8.5,0)}]   (5,-.5)--(-1,-.5) ;
		\node[scale=2] at (2,-1) {$2t+1$};
		\node[scale=2] at (10,-1) {$2t+1$};
	},\label{eq:foldedtrace}
\end{align}
where now the states $\ket{\mcirc}, \ket{\square}$ represent two different contractions in the space of replicas, defined as in \eqref{eq:staggeredstates}. Equation \eqref{eq:foldedtrace} can be further simplified, by noting that the matrix MPS transfer matrix has an eigenvalue fixed by the compatibility condition \eqref{eq:compatibilitycondition}. Using the unitarity of the gate, it is easy to show that\begin{align}
	\fineq[-0.8ex][0.75][1]{
		\MPSinitialstate[0][0][bertiniorange][topright][n]
		\cstate[-1][-.05][]
		\cstate[-.5][.5][]
		\cstate[.5][.5][]
	}=
d^n	\fineq[-0.8ex][0.75][1]{
		\draw (-1,-.05)--++(1,0);
		\cstate[-1][-.05][]
	}\qquad \fineq[-0.8ex][0.75][1]{
	\MPSinitialstate[0][0][bertiniorange][topright][n]
	\cstate[1][-.05][]
	\cstate[-.5][.5][]
	\cstate[.5][.5][]
}=d^n
\fineq[-0.8ex][0.75][1]{
\draw (-1,-.05)--++(-1,0);
\cstate[-1][-.05][]
},
\end{align}
    where, due to the global replica symmetry, the same relation holds if we replace all the circles with squares.
If we assume this eigenvalue of the transfer matrix is nondegenerate and maximal, then we can further simplify \eqref{eq:foldedtrace} as
\begin{align}
\tra{\rho_A^n}=\frac{1}{\chi^n d^{n (L_A+2t+1)}}	\fineq[-0.8ex][0.55][1]{
	\foreach \i in {0,...,6}
	{
		\MPSinitialstate[\i*2][0][bertiniorange][topright][n]
		\ifnumless{\i}{2}{
			\foreach \j in {\i,...,1}
			{
				\roundgate[\j*2+3-\i-2][\i+1][1][topright][bertiniorange][n]
				\roundgate[\j*2+3-\i+6][\i+1][1][topright][bertiniorange][n]
			}
		}
		{}
	}
	\foreach\i in {0,1,2}
	{	
		\cstate[-.5+\i][.5+\i]\cstate[12.5-\i][.5+\i]
	}
	\foreach\i in {0,...,2}
	{	
		\sqrstate[2.5+\i][2.5-\i]
		\sqrstate[9.5-\i][2.5-\i][]
	}
	\sqrstate[5.5][.5]
	\sqrstate[6.5][.5]
	\cstate[-1][-.05]
	\cstate[13][-.05]
	\draw [decorate, decoration = {brace}]   (5,-.5)--(-1,-.5) ;
	\draw [decorate, decoration = {brace},shift={(8.5,0)}]   (5,-.5)--(-1,-.5) ;
	\node[scale=2] at (2,-1) {$2t+1$};
	\node[scale=2] at (10,-1) {$2t+1$};
}.\label{eq:foldedtracesimplified}
\end{align}
Our last step consists into taking the scaling limit $L_A,t\rightarrow\infty$; keeping the ratio ${t}/{L_A}={\rm const}<{1}/{2}$, in such a way that the two edges of the subsystem $A$ are not causally connected.
In this limit, the $L_A-2t-1$ MPS transfer matrices in between the two triangles in \eqref{eq:foldedtracesimplified} can be substituted by the projector on their largest eigenvalue; explicitly we can write 
\begin{align}
	\lim_{m\rightarrow\infty}	\left(\frac{1}{d^n}
	\fineq{
		\MPSinitialstate[0][-1][bertiniorange][topright][n]
		\sqrstate[-.5][-.5][]
		\sqrstate[.5][-.5][]
	}\right)^m=\frac{1}{\chi^n}\fineq{
		\draw(-.5,-.5)--++(-.5,0);
		\draw(.5,-.5)--++(.5,0);	 
		\sqrstate[-.5][-.5][]
		\sqrstate[.5][-.5][]
	}	  \label{eq:Mmatrixfolded},
\end{align}
which allows to simplify Eq.~\eqref{eq:foldedtracesimplified} as
\begin{align}
	\tra[]{\rho_A^n}=\left(\frac{1}{\chi^n d^{ n (2t+1)}}\fineq[-0.8ex][0.7][1]{
		\foreach \i in {0,...,2}
		{
			\MPSinitialstate[\i*2][0][bertiniorange][topright][n]
			\ifnumless{\i}{2}{
				\foreach \j in {\i,...,1}
				{
					\roundgate[\j*2+3-\i-2][\i+1][1][topright][bertiniorange][n]
				}
			}
			{}
		}
		\foreach\i in {0,1,2}
		{	
			\cstate[-.5+\i][.5+\i]
		}
		\foreach\i in {0,...,2}
		{	
			\sqrstate[2.5+\i][2.5-\i]
		}
		\cstate[-1][-.05]
		\sqrstate[5][-.05]
		\draw [decorate, decoration = {brace}]   (5,-.5)--(-1,-.5) ;
		\node[scale=1.4] at (2,-1) {$2t+1$};
	}\right)^2\label{eq:simplifiedtrace},
\end{align}
leading to Eq.~\eqref{eq:factorizedrenyis}.

\section{Spectrum of $\widetilde{U}\widetilde{U}^\dagger$} 
\label{sec:app1}

In this section we want to study the spectrum of $\tilde{U}\tilde{U}^\dagger$, where $\tilde{U}$ is  obtained by a reshuffling of the indexes of the gate $U$, explicitly 
\be
\bra{kl}\tilde{U}\ket{ij}=\bra{lj}{U}\ket{ki}.
\ee
Since it is convenient to use diagrammatic calculus for our calculations here, we begin by introducing a few more useful diagrams. We represent a single gate as  
\begin{align}
	\bra{kl}U\ket{ij}\equiv\fineq
	{
		\roundgate[0][0][1][topright][bertinired][1]
		\node[scale=1] at (-0.6,0.6) {$k$};
		\node[scale=1] at (0.6,0.6) {$l$};
		\node[scale=1] at (-0.6,-0.5) {$i$};
		\node[scale=1] at (0.6,-0.5) {$j$}; 			
	}=\left(\fineq
	{
		\roundgate[0][0][1][topright][bertiniblue][1]
		\node[scale=1] at (-0.6,0.6) {$k$};
		\node[scale=1] at (0.6,0.6) {$l$};
		\node[scale=1] at (-0.6,-0.5) {$i$};
		\node[scale=1] at (0.6,-0.5) {$j$}; 			
	}\right)^*.
\end{align}
Therefore, unitarity and DU2 property \eqref{eq:2ndH} correspond to 
\begin{align}
	\fineq[-0.8ex][0.75][1]
{
	\roundgate[0][0][1][topright][bertinired][1]
	\roundgate[0][-1][1][bottomright][bertiniblue][1]
}=\fineq[-0.8ex][0.75][1]
{
	\roundgate[0][0][1][topright][bertinired][1]
	\roundgate[0][1][1][bottomright][bertiniblue][1]
}=\fineq[-0.8ex][0.75][1]
{
	\draw[thick] (0,0)--++(0,1.5);
	\draw[thick] (0.75,0)--++(0,1.5);
}\qquad 	\fineq[-0.8ex][0.75][1]{
\roundgate[0][1][1][topleft][bertiniblue][1]
\roundgate[1][0][1][topleft][bertiniblue][1]
\roundgate[-1][1][1][topright][bertinired][1]
\roundgate[-2][0][1][topright][bertinired][1]
\draw (-1.5,-.5)--++(2,0);
}=
\fineq[-0.8ex][0.75][1]{
\draw (-.5,1.5)--++(2,0);
\roundgate[1][0][1][topleft][bertiniblue][1]
\roundgate[0][0][1][topright][bertinired][1]
}\quad
\fineq[-0.8ex][0.75][1]{
\roundgate[0][0][1][topleft][bertiniblue][1]	
\roundgate[1][1][1][topleft][bertiniblue][1]
\roundgate[-1][0][1][topright][bertinired][1]
\roundgate[-2][1][1][topright][bertinired][1]	
\draw (-1.5,1.5)--++(2,0);
}=
\fineq[-0.8ex][0.75][1]{
\draw (-.5,-.5)--++(2,0);
\roundgate[1][1][1][topleft][bertiniblue][1]
\roundgate[0][1][1][topright][bertinired][1]	}.
\label{DU2unfolded1}
\end{align}
In fact, DU2 also implies 
\begin{align}
	\fineq[-0.8ex][0.75][1]{
	\roundgate[0][0][1][topleft][bertinired][1]
	\roundgate[1][0][1][topright][bertiniblue][1]
	\roundgate[-1][-1][1][topleft][bertinired][1]
	\roundgate[2][-1][1][topright][bertiniblue][1]
	\draw[thick](-.5,-1.5)--(1.5,-1.5);
}=
\fineq[-0.8ex][0.75][1]{
	\draw[thick] (-.5,1.5)--++(2,0);
	\roundgate[1][0][1][topright][bertiniblue][1]
	\roundgate[0][0][1][topleft][bertinired][1]}
\quad 
\fineq[-0.8ex][0.75][1]{
	\roundgate[0][0][1][topleft][bertinired][1]
	\roundgate[1][0][1][topright][bertiniblue][1]
	\roundgate[-1][1][1][topleft][bertinired][1]
	\roundgate[2][1][1][topright][bertiniblue][1]
	\draw[thick](-.5,1.5)--(1.5,1.5);
}=
\fineq[-0.85ex][0.75][1]{
	\draw[thick] (-.5,-.5)--++(2,0);
	\roundgate[1][1][1][topright][bertiniblue][1]
	\roundgate[0][1][1][topleft][bertinired][1]}.
\label{DU2unfolded2}
\end{align}
The matrix $\tilde{U}\tilde{U}^\dagger$ is clearly Hermitian, so it can be decomposed into orthogonal eigenspaces, with eigenvalues $\Lambda^2$ which are the squared singular values of $\tilde{U}$ with associated projector $P_\Lambda$. 
In formulae, this reads as
\begin{align}
		\widetilde{U}\widetilde{U}^\dagger=\fineq{		\roundgate[0][0][1][topleft][bertinired][1]
		\roundgate[1][0][1][topright][bertiniblue][1]}\equiv \sum_\Lambda \Lambda^2 P_\Lambda \label{eq:Pdef}
\end{align}
Now consider the following quantity, obtained taking the partial trace of $\left(\widetilde{U}\widetilde{U}^\dagger\right)^n$
\begin{align}
&A=	\fineq[-0.8ex][0.7]{
	\foreach\i in {0,...,2}
	{\roundgate[2*\i][0][1][topleft][bertinired][1]
	\roundgate[2*\i+1][0][1][topright][bertiniblue][1]
	}	\hook[-1][1][t][r]
\draw[thick] (5.5,.5) arc (135:-90:0.15);
\draw[thick,dashed] (-.5,.5-.22)--(5.57,.5-.22);
}=\sum_\Lambda \Lambda^{2n} \fineq[-0.8ex][0.7]{
\draw[thick,dashed] (-.5,1.2)--(1.5,1.2);
\Pproj{0.5}{1}
\draw[thick] (-.5,1.5) arc (90:270:0.15);
\draw[thick] (1.5,1.5) arc (90:-90:0.15);
},\label{eq:coso1}
\end{align}
Using Eq.~\eqref{DU2unfolded2} multiple times, we can rewrite Eq.~\eqref{eq:coso1} as
\begin{align}
	A=\frac{1}{d}		\fineq[-0.8ex][0.7][1]{
				\foreach\i in {0,...,2}
			{	\hook[-1][0][b][r]
				\hook[-2][2][t][r]
				\draw[thick] (10.5,1.5) arc (135:-90:0.15);
				\draw[thick,dashed] (-1.5,1.5-.22)--(10.57,1.5-.22);
				\draw[thick] (9.57,-.22) arc (90:-135:0.15);
				\draw[thick,dashed] (-.5,-.22)--(9.57,-.22);
				\roundgate[0+4*\i][0][1][topleft][bertinired][1]
				\roundgate[1+4*\i][0][1][topright][bertiniblue][1]
				\roundgate[-1+4*\i][1][1][topleft][bertinired][1]
				\roundgate[2+4*\i][1][1][topright][bertiniblue][1]
					\draw[thick](-.5+4*\i,1.5)--(1.5+4*\i,1.5);
			}\foreach\i in {0,...,1}
		{	
		\draw[thick](1.5+4*\i,-.5)--(3.5+4*\i,-.5);
	}
		}.
\end{align}
Next, using (the complex conjugate of) Eq.~\eqref{DU2unfolded2} we find 
\begin{align}
&A=\frac{1}{d}		\fineq[-0.8ex][0.64][1]{
	\hook[-1][0][b][r]
	\hook[-2][2][t][r]
\draw[thick] (10.5,1.5) arc (135:-90:0.15);
\draw[thick] (9.57,-.22) arc (90:-135:0.15);
\draw[thick,dashed] (-.5,-.22)--(9.57,-.22);
\draw[thick,dashed] (-1.5,1.5-.22)--(10.57,1.5-.22);
			\roundgate[-1][1][1][topleft][bertinired][1]
\roundgate[10][1][1][topright][bertiniblue][1]
		\foreach\i in {0,...,2}
		{	\roundgate[0+4*\i][0][1][topleft][bertinired][1]
			\roundgate[1+4*\i][0][1][topright][bertiniblue][1]
			}
		\draw[thick](-.5,1.5)--(9.5,1.5);
		\foreach\i in {0,...,1}
		{	
			\draw[thick](1.5+4*\i,-.5)--(3.5+4*\i,-.5);
			\draw[thick](1.5+4*\i,.5)--(3.5+4*\i,.5);
		}
	}=
\frac{1}{d}\fineq[-0.8ex][0.64][1]{	\hook[-1][1][t][r]
	\draw[thick] (9.5,.5) arc (135:-90:0.15);	
	\draw[thick,dashed] (-.5,.5-.22)--(9.57,.5-.22);
	\hook[-1][0][b][r]
	\draw[thick] (9.57,-.22) arc (90:-135:0.15);
	\draw[thick,dashed] (-.5,-.22)--(9.57,-.22);
	\foreach\i in {0,...,2}
	{	\roundgate[0+4*\i][0][1][topleft][bertinired][1]
		\roundgate[1+4*\i][0][1][topright][bertiniblue][1]
	}
	\draw[thick](-1.5,.5)--(-.5,1.5)--(9.5,1.5)--(10.5,.5);
	\foreach\i in {0,...,1}
	{	
		\draw[thick](1.5+4*\i,-.5)--(3.5+4*\i,-.5);
		\draw[thick](1.5+4*\i,.5)--(3.5+4*\i,.5);
	}}=\notag\\
=&\frac{\mathds{1}_d}{d} \sum_\Lambda {\Lambda^{2n}}{} \tra{P_\Lambda},\label{eq:coso2}
\end{align}
where unitarity was used in the last step. Comparing Eqs.~\eqref{eq:coso1} and \eqref{eq:coso2}, we get immediately
\begin{align}
	\fineq[-0.8ex][0.7]{
		\draw[thick,dashed] (-.5,1.2)--(1.5,1.2);
		\Pproj{0.5}{1}
		\draw[thick] (-.5,1.5) arc (90:270:0.15);
		\draw[thick] (1.5,1.5) arc (90:-90:0.15);
	}= \frac{\tra{P_\Lambda}\mathds{1}_d}{d}, \qquad \forall \Lambda\,.
		\label{eq:lemma1}
\end{align}
With a completely analogous reasoning we can also obtain 
\begin{align}
	\fineq[-0.8ex][0.7]{
		\draw[thick,dashed] (-.5,.2)--(1.5,.2);
		\Pproj{0.5}{1}
		\draw[thick] (-.5,.5) arc (90:270:0.15);
		\draw[thick] (1.5,.5) arc (90:-90:0.15);
	}=\frac{\tra{P_\Lambda}\mathds{1}_d}{d}, \qquad \forall \Lambda\,.
	\label{eq:lemma1.1}. 
\end{align}
Now consider 
\begin{align}
&	B=		\fineq[-0.8ex][0.7]{
		\foreach\i in {0,...,3}
		{\roundgate[2*\i][0][1][topleft][bertinired][1]
			\roundgate[2*\i+1][0][1][topright][bertiniblue][1]
		}	\hook[-1][1][t][r]
		\draw[thick] (7.5,.5) arc (135:-90:0.15);
		\draw[thick,dashed] (-.5,.5-.22)--(7.57,.5-.22);
	\hook[-1][0][b][r]
	\draw[thick] (7.57,-.22) arc (90:-135:0.15);
	\draw[thick,dashed] (-.5,-.22)--(7.57,-.22);
}=\sum_\Lambda \Lambda^{2n }\tra{P_\Lambda}\label{eq:coso3}
\end{align}
Using Eqs.~\eqref{DU2unfolded1} $m+1$ times, we get\begin{align}
	B=\frac{1}{d}\fineq[-0.8ex][0.7]{
		\foreach\i in {2,...,3}
		{\roundgate[2*\i][0][1][topleft][bertinired][1]
			\roundgate[2*\i+1][0][1][topright][bertiniblue][1]
		}	\hook[3][1][t][r]
		\hook[3][0][b][r]
			\foreach\i in {1.5,2.5}
		{	\roundgate[4+4*\i][0][1][topleft][bertinired][1]
			\roundgate[5+4*\i][0][1][topright][bertiniblue][1]
			\roundgate[3+4*\i][1][1][topleft][bertinired][1]
			\roundgate[2+4*\i][1][1][topright][bertiniblue][1]
\draw[thick](1.5+4*\i,-.5)--(3.5+4*\i,-.5);
		}
			\draw[thick](5.5+4,1.5)--(3.5+4*2,1.5);			
	\draw[thick,dashed] (3.5,.5-.22)--(15.57,.5-.22);
	\hook[7][2][t][r]
			\draw[thick] (15.5,.5) arc (135:-90:0.15);
	\draw[thick] (13.5,1.5) arc (135:-90:0.15);
	\draw[thick,dashed] (7.3,1.5-.22)--(13.57,1.5-.22);
	\draw[thick] (15.57,-.22) arc (90:-135:0.15);
	\draw[thick,dashed] (3.5,-.22)--(15.57,-.22);
	},
\end{align}
then, using Eq \eqref{DU2unfolded2} $m$ times ($m=1$ in the diagram) we find 
\begin{align}
B=\frac{1}{d}\fineq[-0.8ex][0.7]{
	\foreach\i in {2,...,3}
	{\roundgate[2*\i][0][1][topleft][bertinired][1]
		\roundgate[2*\i+1][0][1][topright][bertiniblue][1]
	}	\hook[3][1][t][r]
	\hook[3][0][b][r]
	\draw[thick,dashed] (3.5,-.22)--(11.57,-.22);
	\foreach\i in {1.5,2.5}
	{	\roundgate[3+4*\i][1][1][topleft][bertinired][1]
		\roundgate[2+4*\i][1][1][topright][bertiniblue][1]		}
				\draw[thick](5.5+4,1.5)--(3.5+4*2,1.5);			
	\draw[thick,dashed] (3.5,.5-.22)--(15.57,.5-.22);
	\hook[3][0][b][r]
	\draw[thick] (15.5,.5) arc (135:-90:0.15);
	\hook[7][2][t][r]
	\draw[thick] (13.5,1.5) arc (135:-90:0.15);
	\draw[thick,dashed] (7.3,1.5-.22)--(13.57,1.5-.22);
	\draw[thick] (15.57,-.22) arc (90:-135:0.15);
	\draw[thick,dashed] (3.5,-.22)--(15.57,-.22);
	\roundgate[4+4*2.5][0][1][topleft][bertinired][1]
	\roundgate[5+4*2.5][0][1][topright][bertiniblue][1]
	\draw[thick](1.5+4*1.5,-.5)--(3.5+4*2.5,-.5);
	\draw[thick](1.5+4*2.5,.5)--(3.5+4*1.5,.5);
},
\end{align}
where now we have $n-m$ blue gates on the bottom and $m+1$ on top. Using Eqs.~\eqref{eq:coso1} and  \eqref{eq:lemma1}, we can simplify the gates on top to write
\begin{align}
	B=\sum_\Lambda \Lambda^{2(m+1)}\frac{\tra{P_\Lambda}}{d^2}\fineq[-0.8ex][0.7]{\foreach\i in {1.5,2,2.5}
	{	\roundgate[3+4*\i][1][1][topleft][bertinired][1]
		\roundgate[2+4*\i][1][1][topright][bertiniblue][1]
		}
	\hook[7][2][t][r]
		\hook[7][1][b][r]
	\draw[thick] (13.5,1.5) arc (135:-90:0.15);
	\draw[thick,dashed] (7.3,1.5-.22)--(13.57,1.5-.22);
		\draw[thick] (13.57,1-.22) arc (90:-135:0.15);
	\draw[thick,dashed] (7.2,1-.22)--(13.57,1-.22);
			}\notag=\\
		=\sum_{\Lambda,\Lambda'}{\Lambda'}^{2(n-m)}{\Lambda}^{2(m+1)} \frac{\tra{P_\Lambda}\tra{P_{\Lambda'}}}{d^2}.
\label{eq:coso4}
\end{align}
Combining Eqs.~\eqref{eq:coso3} and \eqref{eq:coso4}, we get
\begin{align}
	\forall n,m \quad \sum_{\Lambda,\Lambda'}{\Lambda'}^{2(n-m)}{\Lambda}^{2(m+1)} \frac{\tra{P_\Lambda}\tra{P_{\Lambda'}}}{d^2}=\sum_\Lambda \Lambda^{2n} \tra{P_\Lambda}\label{eq:propertytrace}
\end{align}
which implies there can be only one nonzero $\Lambda$. In turn, this implies
\begin{align}
	\tra{P_\Lambda}\Lambda =d^2\qquad 
	\tra{(\widetilde{U}\widetilde{U}^\dagger)^n} =\Lambda^{2(n-1)}d^2\label{eq:spectrumUtilde}
\end{align}
Substituting \eqref{eq:spectrumUtilde} in \eqref{eq:entanglementspeedgeneral} and setting $n_\Lambda=\tra{P_\Lambda}$ this gives Eq.~\eqref{eq:finalformula}.

\section{Parameterisation for Numerical Experiments}
\label{sec:parameterisation}

In order to produce the plots in Fig. \ref{plot:membraneandgrowth}, we used a DU2 gate with $d=2$ and $n_\Lambda=2$, parameterized as follows
\begin{align}
&U=u_0 (v_1 \otimes v_2),
\\
&u_0=
	\begin{pmatrix}
		e^{i\pi/4} &0&0&0\\
		0&0&e^{-i\pi/4}&0\\
		0&e^{-i\pi/4}&0&0\\
		0&0&0&e^{i\pi/4} 	
	\end{pmatrix},\qquad
v_{1/2} = \frac{1}{\sqrt{2}}\begin{pmatrix}
e^{i\alpha_{1/2}} &	-e^{-i\alpha_{1/2}}
	\\
	e^{i\alpha_{1/2}} &	e^{-i\alpha_{1/2}}
\end{pmatrix},
\end{align}
where the $\alpha$s have been fixed to the values $\alpha_1=0.2,\alpha_2=0.3$.
The initial states chosen are random dimer states
\begin{align}
	\ket{\Psi_0}=\left(\sum_{i,j} m_{ij} \ket{i,j}\right)^{\otimes L},
\end{align}
where the matrix $m_{ij}$ is the normalized matrix whose elements, in the computation basis, are pseudo-random numbers distributed uniformly in $[0,1]$.

	%%%%%%%%%%%%%%%%%%%%%%%%%%%%%%%%%%%%%%%%%%%%
	
\end{document}